\documentclass[aps,prd,preprint,groupedaddress]{revtex4}

\usepackage{graphicx}
\usepackage{amsmath}

\begin{document}

\def\be{\begin{equation}}
\def\ee{\end{equation}}

\boldmath
\title{Large $x$ Resummation in $Q^2$ Evolution}
\unboldmath
\author{S. Albino}
\affiliation{{II.} Institut f\"ur Theoretische Physik, Universit\"at Hamburg, Luruper Chaussee 149, 22761 Hamburg, Germany}
\author{B. A. Kniehl}
\affiliation{{II.} Institut f\"ur Theoretische Physik, Universit\"at Hamburg, Luruper Chaussee 149, 22761 Hamburg, Germany}
\author{G. Kramer}
\affiliation{{II.} Institut f\"ur Theoretische Physik, Universit\"at Hamburg, Luruper Chaussee 149, 22761 Hamburg, Germany}
\date{\today}
\begin{abstract}
The standard analytic solution to the Dokshitzer-Gribov-Lipatov-Altarelli-Parisi (DGLAP) equation
in Mellin space is improved by resumming the large $x$ divergences. 
Explicit results are given to next-to-leading order and next-to-leading logarithmic accuracy.
Numerically, the theoretical error is found to be reduced by the resummation 
for a large range of $x$.
\end{abstract}


\maketitle



Parton distribution functions (PDFs) are indispensable for the perturbative calculation of all high energy processes
involving incoming hadrons. In particular, their precise knowledge is essential for the 
successful interpretation of the LHC experiments. Likewise, fragmentation functions (FFs)
are required for inclusive hadron production calculations. By solving the spacelike/timelike DGLAP equation \cite{Gribov:1972ri}
\be
\frac{d}{d\ln Q^2} D(N,Q^2)=P(N,a_s(Q^2)) D(N,Q^2)
\label{dglap}
\ee
for the vector $D$ of PDFs/FFs (the formulae of this letter apply to either causality, which is therefore not specified), 
their $Q^2$ evolution is determined since the matrix of splitting
functions $P$ can be calculated perturbatively as $P=\sum_{n=1}^\infty a_s^n P^{(n-1)}$,
where $a_s=\alpha_s/(2\pi)$, $\alpha_s$ being the strong coupling. As $P$
reaches next-next-to-leading order (NNLO) precision \cite{Moch:2004pa}, resummation of
divergences at phase space extremes becomes relevant. This letter is concerned with
soft gluon divergences at large $x$, which have a general relevance since 
their effect propagates to all $x$ through the evolution. For example, we show later that, relative to standard NLO evolution,
the reduction in the theoretical uncertainty due to large $x$ resummation
is comparable to that on going to NNLO for a large range of $x$. In general,
these divergences take the form $a_s^n [\ln^{n-r} (1-x)/(1-x)]_+$ (or $a_s^n \ln^{n+1-r} N$ in Mellin space), 
and $r=0,...,n$ labels the class of divergence. They factor out of a hard subprocess
according to \cite{Contopanagos:1996nh}
\be
W(a_s,N)=W_{\rm res}(a_s,N)\left(\sum_n a_s^n W_{\rm FO}^{(n)}(N)\right),
\label{genresproc}
\ee
where the fixed order (FO) series, in parenthesis, is free of
these divergences since they are all contained in $W_{\rm res}$. At large $N$, 
$W$ is approximated by $W_{\rm res}$ if its divergences are resummed, which involves
writing it as an exponential, then, in the exponent, grouping terms of the same class,
giving a series in $a_s$ keeping $a_s \ln N$ fixed.
Such expansions exist for coefficient functions for various processes, e.g.\
inclusive hadron production \cite{Cacciari:2001cw}, deeply inelastic scattering and the Drell-Yan process
\cite{Catani:1989ne}. 

In this letter we present an analytic solution to the DGLAP equation
which improves the FO accuracy by resumming large $x$ divergences. 
We work in a minimal subtraction (MS) scheme where $P$ is resummed:
its diagonal components approach
$[1/(1-x)]_+$ ($\ln N$) while the off-diagonal components become constant \cite{Korchemsky:1988si}.
Our approach is based on the commonly used Mellin space solution \cite{Furmanski:1981cw}:
The DGLAP evolution is written in the form
\be
D(N,Q^2)=E(N,a_s(Q^2),a_s(Q_0^2))D(N,Q_0^2),
\label{solofdglap}
\ee
then the $a_s=a_s(Q^2)$ and $a_0=a_s(Q_0^2)$ dependences in the higher order parts of $E$ are factored out,
\be
E(N,a_s,a_0)=U(N,a_s)E_{\rm LO}(N,a_s,a_0)U^{-1}(N,a_0),
\label{waytosolvedglap}
\ee
where $E_{\rm LO}$ is the LO evolution, given formally by
\be
E_{\rm LO}(N,a_s,a_0)=\exp\left[-\frac{P^{(0)}(N)}{\beta_0}\ln \frac{a_s}{a_0}\right].
\label{formalresforELO}
\ee
Here, $\beta_0$ is the first coefficient appearing in the perturbative expansion for the evolution of $a_s$,
$da_s(\mu^2)/d\ln \mu^2=-\sum_{n=0}^\infty \beta_n a_s^{n+2}(\mu^2)$. Then 
\be
\frac{dU}{da_s}=-\frac{R}{\beta_0}+\frac{1}{\beta_0 a_s}\left[U,P^{(0)}\right]
\label{UfromP2}
\ee
replaces the DGLAP equation, where 
\be
R=\sum_{n=1}^\infty a_s^{n-1} R^{(n)}=-\beta_0\left(\frac{P}{\beta}+\frac{P^{(0)}}{\beta_0 a_s}\right)U.
\label{defofr}
\ee
The accuracy of the evolution depends on how the matrix $U$ is approximated. The standard approach is to use the
the FO series
\be
U(N,a_s)={\mathbf 1}+\sum_{n=1}^\infty U^{(n)}(N)a_s^n,
\label{expanU}
\ee
which can be determined order by order from Eq.\ (\ref{UfromP2})
since the $R^{(n)}$ depend on the $U^{(m)}$ for $m\le n-1$ only. However,
the accuracy may be improved by resumming the large $x$ divergences, which is the main task of this letter.
The result is that at large $N$, its diagonal components take the form
$U_{II}(N,a_s)\simeq \exp [\sum_{r=1}^\infty U^{[r]}_{II}(a_s \ln N)a_s^{r-1}]$
(recall that $r$, in the square brackets here, labels the class of divergence).

The resummation of the non singlet (or valence quark) component of $E$
is trivially implemented by solving the DGLAP equation exactly. However, we discuss it in detail
since the resummation in the singlet evolution to be performed later is similar. 
Equation (\ref{waytosolvedglap}) simplifies to the scalar equation
$E_{\rm NS}=\left(a_s/a_0\right)^{-P^{(0)}_{\rm NS}/\beta_0} (U_{\rm NS}(a_s)/U_{\rm NS}(a_0))$.
In the unresummed case, $U_{\rm NS}(a_s)/U_{\rm NS}(a_0)$ is expanded as a series in $a_s,a_0$, e.g.\ at NLO 
it is taken as $1-(a_s-a_0)R^{(1)}_{\rm NS}/\beta_0$, where $R^{(1)}_{\rm NS}=P^{(1)}_{\rm NS}-(\beta_1/\beta_0)P^{(0)}_{\rm NS}$. 
In the resummed case, it is calculated in the form given by Eq.\ (\ref{genresproc}), e.g.\ at NLO it becomes
$(U_{\rm NS}^{[1]}(a_s)/U_{\rm NS}^{[1]}(a_0))\left[1-(a_s-a_0)
\left(U_{\rm NS}^{[1](1)}+R^{(1)}_{\rm NS}/\beta_0\right)\right]$, where $U_{\rm NS}^{[1](1)}$ is the coefficient
of the $O(a_s)$ term in the FO expansion of $U_{\rm NS}^{[1]}$. To obtain the $U_{\rm NS}^{[r]}$ requires solving the
equation $dU_{\rm NS}/da_s=Q_{\rm NS}U_{\rm NS}$ at large $N$, 
where $Q_{\rm NS}=P_{\rm NS}/\beta+P^{(0)}_{\rm NS}/(\beta_0 a_s)$. Of course this holds
for all $N$, as can be seen by substituting Eq.\ (\ref{waytosolvedglap}) 
into the DGLAP equation. For example, at NLO, the next-to-leading logarithms (NLLs), being $r=1$ ($O(a_s \ln N)$) terms, 
are resummed by taking $U_{\rm NS}^{[1]}=\exp [a_s U_{\rm NS}^{[1](1)}]$
with $U_{\rm NS}^{[1](1)}=-R^{(1)}_{\rm NS}/\beta_0+O(1)$, and taking the last equation to be exact leads to
the exact solution to the NLO DGLAP equation, $E_{\rm NS}=(a_s/a_0)^{-P^{(0)}_{\rm NS}/\beta_0}
\exp[-(a_s-a_0)R^{(1)}_{\rm NS}/\beta_0]$.

Analytic resummation of the singlet $E$ is nontrivial, since its matrix
structure rules out an exact and analytic solution to the DGLAP equation . We will present
for the first time an analytic approach
for resumming the FO singlet $E$. We first derive in detail the unresummed $E$, which will be modified
later in order to implement the resummation.
To put $E_{\rm LO}$ in Eq.\ (\ref{formalresforELO}) into a form which 
can be explicitly evaluated, we diagonalize $P^{(0)}$,
\be
P^{(0)}=\lambda_+ M^+ +\lambda_- M^-,
\label{P0fromMlam}
\ee
where 
\be
\lambda_\pm=\frac{1}{2}\left[P_{qq}^{(0)}+P_{gg}^{(0)}\pm \sqrt{(P_{qq}^{(0)}-P_{gg}^{(0)})^2-4 P_{qg}^{(0)} P_{gq}^{(0)}}\right],
\label{defoflam}
\ee
are the eigenvalues of $P^{(0)}$, and the projection operators 
\be
M^\pm =\frac{1}{\lambda_\pm -\lambda_\mp} \left[P^{(0)}-\lambda_\mp {\mathbf 1}\right]
\ee
obey $M^\pm M^\mp=0$, $M^\pm M^\pm= M^\pm$, $\sum_i M^i ={\mathbf 1}$.
Then 
\be
E_{\rm LO}(N,a_s,a_0)=\sum_i 
M^i(N)\left(\frac{a_s}{a_0}\right)^{-\frac{\lambda_i(N)}{\beta_0}}.
\ee

To diagonalize all occurrences of $P^{(0)}$ in $U$,
we work with the $ij$th ``components'' projected out 
by operating on the left by $M^i$ and on the right by $M^j$.
Note that the sum of all ``components'' of any $2\times 2$ 
matrix $A$ gives back the full result, i.e.\ $A=\sum_{ij} M^i A M^j$. 
Matching coefficients of $a_s$ in Eq.\ (\ref{UfromP2}) gives
\be
U^{(n)}=\sum_{ij} \frac{1}{\lambda_j -\lambda_i -\beta_0 n} M^i R^{(n)} M^j.
\label{coeffsinU}
\ee
(Note that $P^{(0)}M^i=M^i P^{(0)}=\lambda_i M^i$, which follows from Eq.\ (\ref{P0fromMlam}), has been used here.)
The right hand side is clearly a sum over the ``components'' $M^i U^{(n)} M^j$.

At NLO,
\be
U=\sum_{ij} M^i \left[{\mathbf 1}+\frac{1}{\lambda_j -\lambda_i -\beta_0}a_s R^{(1)}\right] M^j,
\label{genexpforUNLO}
\ee
where
\be
R^{(1)}=P^{(1)}-\frac{\beta_1}{\beta_0}P^{(0)}.
\label{R1fromP1P0}
\ee
In $U$, and therefore in general in $E$, there is a simple pole on the positive real axis of $N$ space
when $\lambda_+-\lambda_- -\beta_0=0$ and 2 simple and 2 double poles off the real axis when $\lambda_+-\lambda_- =0$,
which will limit the choice of the contour used for the inverse Mellin transform.
Fortunately, they manifest themselves in Eq.\ (\ref{waytosolvedglap}) as spurious NNLO terms,
which may be subtracted. The NLO expansion in $a_s,a_0$ with $E_{\rm LO}$ fixed,
\be
E=E_{\rm LO}+U^{(1)}E_{\rm LO}a_s-E_{\rm LO}U^{(1)}a_0,
\label{Ewoshoterms}
\ee
is free of these poles,
which is most easily seen in its projections: The ``diagonal'' ($i=j$) components read
\be
M^i E M^i = \left(\frac{a_s}{a_0}\right)^{-\frac{\lambda_i}{\beta_0}}M^i\left[{\mathbf 1}  -(a_s-a_0)
\frac{R^{(1)}}{\beta_0}\right]M^i,
\label{unresdiag}
\ee
and the ``off-diagonal'' ($i\neq j$) components read
\be
M^i E M^j =\frac{1}{\lambda_j-\lambda_i -\beta_0}
\left[\left(\frac{a_s}{a_0}\right)^{-\frac{\lambda_j}{\beta_0}}a_s
-\left(\frac{a_s}{a_0}\right)^{-\frac{\lambda_i}{\beta_0}}a_0\right]
M^i R^{(1)} M^j.
\label{offdiagpartinres}
\ee
It is clear that the pole in $U$ for which $\lambda_j-\lambda_i -\beta_0=0$ only appears in the
``off-diagonal'' component $M^i E M^j$. Its cancellation can be seen by
setting $\lambda_j=\lambda_i +\beta_0+\epsilon$ and taking the limit $\epsilon\rightarrow 0$.
The components of $U$ containing simple and double poles for which
$\lambda_+-\lambda_- =0$ appear in both the ``off-diagonal'' and ``diagonal'' components of $E$.
The $1/(\lambda_+ -\lambda_-)$ terms cancel since $E$ is invariant with respect to the interchange
$\lambda_+ \leftrightarrow \lambda_-$. The $1/(\lambda_+ -\lambda_-)^2$ terms, 
which give rise to both simple and double poles, can be seen to cancel by taking
$\lambda_+ =\lambda_-+\epsilon$ etc.\ as before.

Now we incorporate the resummation in the solution. 
We require the asymptotic behaviour of $U$ at large $N$. This is determined by using
Eq.\ (\ref{UfromP2}), or equivalently (\ref{coeffsinU}), to obtain
the $U^{(n)}$ for all $n$ as functions of the $P^{(m)}$, $m=0,...,n$, whose asymptotic behaviour in MS schemes 
was given in the quark-gluon basis earlier. 
Fortunately, this basis' components coincide with the projections we have been using: From Eq.\ (\ref{defoflam}),
the large $N$ behaviour of $P$ implies that $\lambda_{+(-)} \approx P_{qq(gg)}^{(0)}$, therefore
\begin{eqnarray}
M^{+(-)}\approx &\left( \begin{array}{cc}
1(0) & 0 \\
0 & 0(1)
\end{array} \right)+O\left(\frac{1}{\ln N}\right).
\end{eqnarray}
This means that performing the projection $M^i A M^j$ has the same effect as setting all components
to zero except $A_{\alpha_i \alpha_j}$, where $\alpha_{+(-)}=q(g)$. One immediately finds from Eq.\ (\ref{coeffsinU})
that the ``off-diagonal'' components of $U$ fall like $1/\ln N$ and may therefore 
be left as a series in $a_s$, while
the ``diagonal'' components of $U$ cannot be approximated in this way since 
$M^i U^{(n)} M^i$ grows like $\ln^n N$. To resum it, we solve the
``diagonal'' components of Eq.\ (\ref{UfromP2}) in the large $N$ limit. The matrix multiplications on the right hand side
can be decomposed according to
$AB=\sum_{ijk}(M^i A M^j) (M^j B M^k)$, where $B$ is also any 2$\times$2 matrix. Then any product of an
``off-diagonal'' component of $U$ with an ``off-diagonal'' component of $R$ or $P^{(0)}$ in either order 
contains no divergences and thus is neglected. Neglecting $O(1)$ terms, the result is
\be
\frac{d}{da_s} (M^i U M^i)=Q_{ii} (M^i U M^i),
\label{resumdiagU}
\ee
where the matrix $Q_{ii}=M^i \left(P/\beta+P^{(0)}/(\beta_0 a_s)\right) M^i$ at large $N$.
This can be solved order by order to obtain a series for $M^i U M^i$ which can be identified and resummed. 
The diagonal component $(Q_{ii})_{\alpha_i \alpha_i}$ grows like
$\ln N$, while the other diagonal component falls like $1/\ln N$. The off-diagonal components both approach a constant.

The NLO evolution contains divergences up to the NLL level.
The $r=0$ (LL) terms are all contained in $E_{\rm LO}$.
The $r=1$ terms are all contained in $U$, and according to Eq.\ (\ref{resumdiagU}) can
be accounted for by taking
\be
M^i U M^i= U_i^{[1]}M^i+O(a_s(a_s \ln N)^m),
\ee
where the $U_i^{[1]}$ are constrained at large $N$ by writing
\be
U_i^{[1]}=\exp\left[a_s U_i^{[1](1)}\right]
\ee
and requiring that
\be
U_i^{[1](1)}= -\frac{1}{\beta_0}R^{(1)}_{\alpha_i \alpha_i } +O(1).
\label{eqforUreq1fromR1}
\ee
The $O(1)$ terms in $U^{[1] (1)}_{ii}$ are not constrained, and will be chosen later. In this approximation,
$R^{(1)}_{\alpha_i \alpha_i}$ is calculated 
with $P_{\alpha_i \alpha_i}^{(n)}\approx \hat{P}_{\alpha_i \alpha_i}^{(n)}\ln N$
in Eq.\ (\ref{R1fromP1P0}), where
\cite{Floratos:1981hs}
\be
\begin{split}
\hat{P}_{qq}^{(0)}= &-2C_F, \\
\hat{P}_{qq}^{(1)}= &C_F\left[C_A \left(\frac{\pi^2}{3}-\frac{67}{9}\right)+\frac{20}{9}T_R n_f \right], \\
\hat{P}_{gg}^{(0,1)}= &\frac{C_A}{C_F}P_{qq}^{(0,1)}.
\end{split}
\label{eqforcoefslnn}
\ee
From these results, and using the form in Eq.\ (\ref{genresproc}), the LL and NLL resummed $U$ at NLO will be chosen as
\be
U=\sum_i U_i^{[1]} M^i\left[{\mathbf 1}
+a_s  \left(U^{(1)}-U_i^{[1](1)}{\mathbf 1}\right)\right] M^i
+a_s\sum_{i\neq j} M^i U^{(1)} M^j.
\label{resumedU}
\ee
The first line gives the resummed form for the ``diagonal'' components, while the ``off-diagonal'' 
components contained in the last line are as for no resummation.
This agrees with the FO series when expanded to NLO, and 
gives the correct large $N$ behaviour, noting that the $\ln N$ divergence
in $M^iU^{(1)}M^i$ cancels that in $U_i^{[1](1)}M^i$.  
Analogously, $U^{-1}$ is found to be equal to Eq.\ (\ref{resumedU})
with $U^{(1)}$ and $U_i^{[1](1)}$ everywhere multiplied by $-1$.
Then $U U^{-1}={\mathbf 1}+O(a_s^2)+O(a_s(a_s \ln N)^m)$, as required. As in the unresummed case,
the resummed $U$ and thus $E$ contains problematic poles at $\lambda_+-\lambda_- -\beta_0=0$ and $\lambda_+-\lambda_- =0$.
To deal with this problem, and remain close to the unresummed approach of Ref.\ \cite{Furmanski:1981cw}, we
expand $E$. Note however that
$U_i^{[1]}$ must be kept fixed to preserve the resummation. Then the ``diagonal'' components read
\be
M^i E M^i =\frac{U_i^{[1]}(a_s)}{U_i^{[1]}(a_0)}\left(\frac{a_s}{a_0}\right)^{-\frac{\lambda_i}{\beta_0}}
M^i \left[{\mathbf 1}
-(a_s-a_0)\left(\frac{R^{(1)}}{\beta_0} +U_i^{[1](1)}{\mathbf 1}\right) \right]M^i
\label{mainresforresinevol}
\ee 
while the ``off-diagonal'' components read
\be
M^i E M^j =\frac{1}{\lambda_j-\lambda_i -\beta_0}
\left[a_s\left(\frac{a_s}{a_0}\right)^{-\frac{\lambda_j}{\beta_0}}\frac{1}{U_j^{[1]}(a_0)}
-a_0\left(\frac{a_s}{a_0}\right)^{-\frac{\lambda_i}{\beta_0}}U_i^{[1]}(a_s)\right]M^i R^{(1)} M^j.
\label{offdiagres}
\ee
These last two equations, together with Eq.\ (\ref{eqforUreq1fromR1}), are our main results.
They agree with Eq.\ (\ref{Ewoshoterms}) when expanded to NLO without keeping
$U_i^{[1]}$ fixed, i.e.\ when the resummation is undone, and obey the normalization condition 
\be
E(N,a_s,a_s)= {\mathbf 1}.
\label{boundcondonE}
\ee

The next step in eliminating these problematic poles involves 
tuning the $U_i^{[1](1)}$ at finite $N$. The infinite number of scheme choices reflects
the uncertainty of the unknown higher order terms, a
ubiquitous feature of perturbation theory. The pole cancellation is easier to achieve by
imposing ``common sense'' constraints obeyed
by the unresummed evolution, e.g.\ the requirements that
$U_i^{[1](1)}=0$ at each pole so that
the resummed evolution coincides with the unresummed one there, that
the $N$ dependence of $U$ occurs only through the $\lambda_\pm$ and $R^{(1)}$, and that
the evolution is invariant under the interchange $\lambda_+ \leftrightarrow \lambda_-$.
Of course, there is also the resummation condition of
Eq.\ (\ref{eqforUreq1fromR1}). One possibility that enforces all these conditions and leads to the pole cancellation is
\be
U_{\pm \pm}^{[1](1)}=K \lambda_\pm \frac{(\lambda_+-\lambda_-)^2\left[(\lambda_+-\lambda_-)^2-\beta_0^2\right]}
{(\lambda_+ +\lambda_-)^4},
\label{finchoiceforU}
\ee
where, for either $i=\pm$ (defining $(C_+,C_-)=(C_F,C_A)$),
\be
K=\frac{1}{2C_i \beta_0}\left(\frac{C_F+C_A}{C_A-C_F}\right)^4
\left(\hat{P}_{\alpha_i \alpha_i}^{(1)}-\frac{\beta_1}{\beta_0}\hat{P}_{\alpha_i \alpha_i}^{(0)}\right).
\ee

\begin{figure}[h!]
\includegraphics[width=12cm,angle=-90]{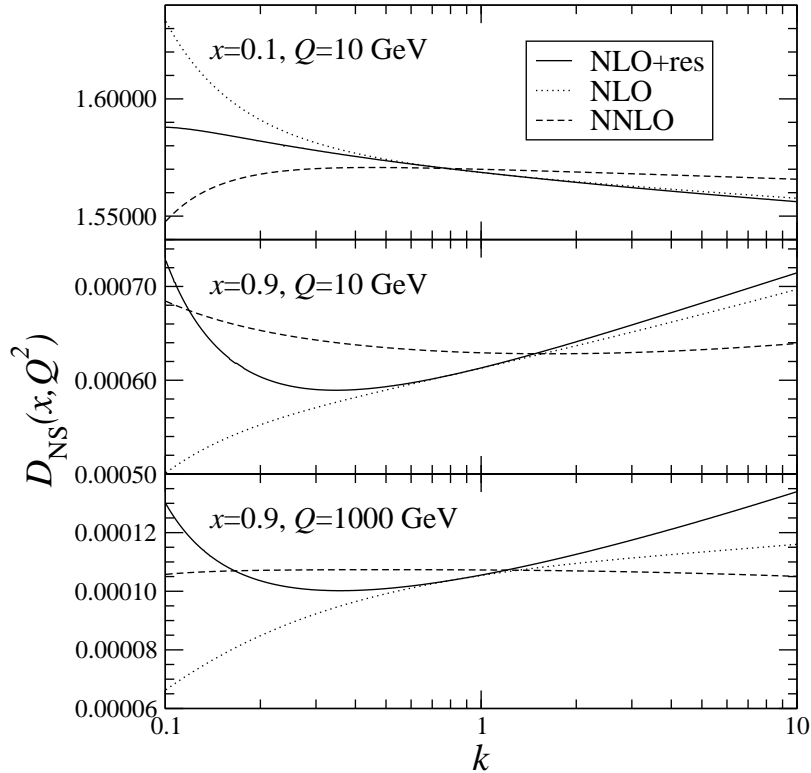}
\caption{\label{fig1} Renormalization scale variation of the non singlet evolved with our resummed NLO approach (solid line), 
as well as with the usual unresummed NLO (dotted) and NNLO (dashed) approaches.}
\end{figure}
Finally, to study the effect of resummation on the theoretical error we
vary the renormalization scale $\mu$ by fixing $\mu^2=k Q^2$ and varying $k$.
We also set $\mu_0^2=k Q_0^2$ in $a_0$ as required by Eq.\ (\ref{boundcondonE}).
This means that $a_{s(0)}=a_s(k Q_{(0)}^2)$ are the new expansion
variables instead of $a_s(Q_{(0)}^2)$, and $(a_s/a_0)^{-\lambda_i/\beta_0}$ and $U^{[1]}_i(a_{s(0)})$ are treated as fixed. 
The NLO relation $a_s(Q_{(0)}^2)=a_{s(0)} (1+a_{s(0)}\beta_0 \ln k)$ implies the replacement
$P^{(1)}\rightarrow P^{(1)}+P^{(0)} \beta_0 \ln k$ everywhere. Note that $\hat{P}_{\alpha_i \alpha_i}^{(1)}
\rightarrow \hat{P}_{\alpha_i \alpha_i}^{(1)} -2C_i \beta_0 \ln k$ in Eq.\ (\ref{eqforcoefslnn}).

For our numerical analysis,
we fix the number of active quark flavours $n_f=5$, put $\Lambda^{(5)}=200$ MeV and $Q_0=2$ GeV, and apply spacelike
evolution to the non singlet distribution 
$D_{\rm NS}(x,Q_0^2)\propto (1-x)^3$, being a typical large $x$ behaviour. Resummation
reduces the scale variation over a wide range of $x$ (see Fig.\ \ref{fig1}), as anticipated in the introduction.
Away from the large $x$ region, the scale variation is even reduced to around 
that of the NNLO calculation \cite{Moch:2004pa}, for low
$Q^2$ values in particular. For further analysis we refer the reader to a 
future publication where also resummation at NNLO will be studied. Fortunately, the three calculations accidentally
give similar results around the commonly chosen value $k=1$, so that unresummed divergences should not have harmed
global fits of PDFs/FFs.

In conclusion, we have presented an analytic approach for resumming large $x$ divergences
in $Q^2$ evolution. This makes a substantial numerical difference to the evolution away from $k=1$.
Since we perform the resummation in the ``diagonalization'' scheme of Ref.\ \cite{Furmanski:1981cw}
that is used to solve the FO DGLAP equation, our approach is the simplest one.
Since it provides both an important conceptual improvement to 
perturbative QCD and leads to a more accurate determination of PDFs and FFs,
our approach is valuable for the description of phenomenology at the LHC and other colliders of the foreseeable future.

This work was supported in part by the German Federal Ministry for
Education and Research BMBF through Grant No. 05 HT6GUA.



\end{document}